# Car following behavioral stochasticity analysis and modelling: Perspective from wave travel time

Junfang Tian[1], Chenqiang Zhu[1], Danjue Chen[2], Rui Jiang[3*], Guanying Wang[1*], Ziyou Gao[3]

[1]Institute of Systems Engineering, College of Management and Economics, Tianjin University, Tianjin 300072, China

[2]Department of Civil and Environmental Engineering, University of Massachusetts Lowell, United States

[3]Key Laboratory of Transport Industry of Big Data Application Technologies for Comprehensive Transport, Ministry of Transport, Beijing Jiaotong University, Beijing 100044, China

This paper analyzes the car following behavioral stochasticity based on two sets of field experimental trajectory data by measuring the wave travel time series $\tilde{\tau}_n(t)$ of vehicle *n*. The analysis shows that (i) No matter the speed of leading vehicle oscillates significantly or slightly, $\tilde{\tau}_n(t)$ might change significantly; (ii) A follower's $\tilde{\tau}_n(t)$ can vary from run to run even the leader travels at the same stable speed; (iii) Sometimes, even if the leader speed fluctuates significantly, the follower can keep a nearly constant value of $\tilde{\tau}_n(t)$. The Augmented Dickey-Fuller test indicates that the time series $\xi_n(t) = d\tilde{\tau}_n(t)/dt$ follows a mean reversion process, no matter the oscillations fully developed or not. Based on the finding, a simple stochastic Newell model is proposed. The concave growth pattern of traffic oscillations has been derived analytically. Furthermore, simulation results demonstrate that the new model well captures both macroscopic characteristic of traffic flow evolution and microscopic characteristic of car following.

# 1 Introduction

Traffic congestion is a nuisance to motorists since it causes more oscillations and discomfort, and results in more fuel consumption and more accidents. Due to the complicated driving behaviors, traffic flow exhibits many fascinating phenomena such as traffic breakdown and capacity drop (Chen et al., 2012a), hysteresis (Ahn, et al., 2013; Chen et al., 2012b), spontaneous formation of jams (Treiterer and Myers, 1974; Zheng et al., 2011), wide scattering of flow density data in congested flow (Treiber et al., 2003). To understand the mechanism of these phenomena is an essential issue for better controlling traffic flow and managing traffic congestion. However, up to now our understanding of traffic flow is limited due to human behavior involved. Actually, there are controversaries (Kerner, 2004; Treiber et al., 2010; Schönhof and Helbing, 2009) in the field. For example, Kerner (2004) have criticized that classical car following models incorporating the fundamental diagram in the steady states cannot reproduce the empirical consistent capacity drop phenomenon. Meanwhile Treiber et al. (2010) and Schönhof and Helbing (2009) have doubted existence of the synchronized traffic flow.

Understanding the driving behavior is also important in the era of connected and automatous vehicles (CAV). Firstly, it would be a long transition towards 100% deployment of CAVs (Mahmassani, 2016). One can expect the new mixed traffic flow of regular vehicles and CAVs for a long time. Secondly, car following data-based modelling approach has been widely applied to control the movements of the automatous vehicles (see, e.g., Zhu et al. 2018). Therefore, clarifying the driving behaviors helps to improve the performance of CAVs.

To simulate and understand the car-following behavior, many car-following models have been proposed. The earliest one might be traced back to the Pipes model (Pipes, 1953), which is based on the assumption that drivers maintain a constant space headway. Since then, many classical models have been developed, to name a few, General Motor models (GM, Gazis et al., 1961), Gipps models (Gipps, 1981), Optimal Velocity model (OVM, Bando et al, 1995), Full Velocity Difference model (FVDM, Jiang et al, 2001), Intelligent Driver model (IDM, Treiber et al, 2000). Some models have been successfully used in commercial software, such as Wiedemann model in VISSIM (Wiedemann, 1974), Fritzsche model in PARAMICS (Fritzsche, 1994), and Gipps model in AIMSUN.

In 2002, Newell proposed a simple model, in which it is assumed that a follower's trajectory overlaps its leader's shifted by a time $\tau$ and spacing $\delta$. In Newell model, traffic oscillations neither amplify nor dampen, which is not consistent with observations. To overcome the deficiency of Newell model, Laval and Leclercq (2010) (LL) proposed that vehicle trajectories accord well with Newell's car-following model before they experience traffic oscillations. However, the trajectories can deviate from Newell trajectories when experiencing traffic oscillations. As a result, the LL model is shown to be able to reproduce traffic oscillations observed in NGSIM data. Later, Chen et al. (2012a) further considered the timid and aggressive driving behaviors in LL model.

The above-mentioned models are deterministic ones. In the models, the growth of traffic oscillations is due to instability of the models themselves. However, recently, more and more studies indicate that stochasticity plays a nontrivial role in traffic flow. Wagner (2012) introduced the strong fluctuation of preferred headway to simulate the macroscopically observable randomness in traffic flow, i.e. the large fluctuations of the spacing around the equilibrium value with zero speed-difference in the car following process. Laval et al. (2014) introduced white noise into the Newell model to simulate acceleration process in free traffic flow, which then enables the model to reproduce the formation of oscillations. Later, Yuan et al. (2018) and Xu and Laval (2019) further improve the model by Laval et al. (2014). Jiang et al. (2014, 2015, 2018) and Huang et al. (2018) have conducted the experimental study of car following behaviors on an open road section. The concave growth pattern of traffic oscillations was discovered, i.e. the standard deviation of vehicle speed grows along the vehicle platoon in a concave way. However, the oscillations initially grow in a convex way in the classical deterministic car-following models mentioned above (such as the GM models, Gipps model, OVM etc., see Jiang et al. (2014, 2015)), which contradicts with the experimental finding.

Theoretically, Treiber and Kesting (2018), Wang et al. (2019), Ngoduy et al. (2019) have studied stochastic linear stability conditions which, capture the effect of stochasticity on traffic flow. The stability results conform to the empirical results that oscillations grow in concave way and traffic instability is related to the stochastic nature of traffic flow.

In this paper, we analyze the car following behavioral stochasticity based on two sets of field experimental trajectory data. Following the framework of Laval and Leclerq (2010), we measure the wave travel time series $\tilde{\tau}_n(t)$ of vehicle *n*. The contributions of the paper are at least twofold. (i) The characteristics of wave travel time $\tilde{\tau}_n(t)$ and its changing rate $\xi_n(t)=d\tilde{\tau}_n(t)/dt$ are analyzed and modelled, which is important not only for microscopic car following behavioral analysis but also for macroscopic traffic flow analysis, such as the stochastic fundamental diagram. The most significant finding is that car following stochasticity, reflecting by $\xi_n(t)$, follows a mean reversion process. (ii) A simple stochastic Newell model is proposed. The analytical results and simulations show that not only the concave growth pattern of traffic oscillations but also the empirical observed congested traffic flow can be reproduced. Furthermore, vehicle trajectories calibrations illustrate that microscopic characteristics in car following process can be simulated very well.

The rest of the paper is organized as follows. Section 2 introduces the study framework. Section 3 analyzes the first set of field experiment and shows that $\xi_n(t)$ follows the mean reversion process. Section 4 analyzes the second set of field experiment that mimics the fully developed oscillations, and confirms the finding in section 3. Section 5 proposes a new stochastic Newell model. The simulation results are given in Section 6. Section 7 demonstrates that stochastic factors might exhibit common feature from different perspectives. Finally, the conclusion is given in Section 8.

## 2 The framework

Newell's car following model (abbreviated as NCM, see Newell, 2002) assumes that the following vehicle's trajectory is the trajectory of the leading vehicle with a translation in time and space. Let $n$-1 denote the vehicle ahead of vehicle $n$, and their location and speed are $x_{n-1}$ and $x_n$, $v_{n-1}$ and $v_n$, respectively. When vehicle $n$-1 changes its speed from $v_{n-1} = v$ to $v_{n-1} = v'$, vehicle $n$ will adjust its speed in the same way after a space displacement of $s_0$ and an adjusting time of $\tau$ to reach the preferred spacing for the new speed $v'$. Under the car-following context, NCM is described by

$$x_n(t+\tau) = x_n(t) + \min\left(x_n^{\text{free}}(t), x_n^{\text{cong}}(t)\right) \tag{1}$$

where the free moving and congested moving terms are defined as

$$x_n^{\text{free}}(t) = v_{\max}\tau \tag{2}$$

$$x_n^{\text{cong}}(t) = x_{n-1}(t) - x_n(t) - \delta \tag{3}$$

where $v_{\max}$ is the maximum speed. NCM adopts a single wave speed $-w = -\delta/\tau = -(L_{veh} + s_0)/\tau$ independent of traffic states. Here $L_{veh}$ is vehicle length. Traffic information, such as speeds, flows, spacing, etc., propagate unchanged along characteristic travelling wave with speed either $v_{\max}$ or $-w$.

However, disturbances will never grow or decay in NCM. Laval and Leclerq (2010) reported that vehicle trajectories will deviate from the Newell trajectories (which are generated by the NCM) when the vehicles go through traffic oscillations. To capture these deviations, the congested moving term is revised as (Chen et al., 2012a),

$$x_n^{\text{cong}}(t) = x_{n-1}(t - \eta_n(t)) - x_n(t) - \delta\eta_n(t) \tag{4}$$

$$\eta_n(t) = \tilde{\tau}_n(t)/\tau \tag{5}$$

The term $\tilde{\tau}_n(t)$ is the actual wave travel time, see Figure 1.

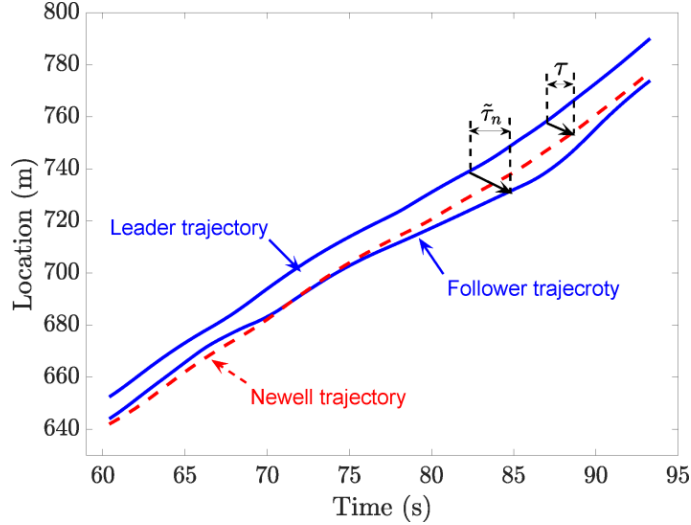

Figure 1. The measurement of $\tilde{\tau}_n(t)$.

Laval and Leclerq (2010) assume that: (i) $\eta_n(t) = 1$ under the equilibrium driving states; (ii) there are three kinds of variation patterns of $\eta_n(t)$ when the vehicles are under the nonequilibrium driving states, i.e. the concave triangle, the convex triangle and constant patterns, see Figure 2, where $\eta_n^0 = \eta_n^1 = 1$ and $\varepsilon_n^0 = \varepsilon_n^1$.

Later, Chen et al. (2012a) have examined the traffic oscillations in the NGSIM US101 trajectory data to verify above assumptions of Laval and Leclerq (2010) and found that: (1) $\eta_n^0(\varepsilon_n^0)$ may be different from $\eta_n^1(\varepsilon_n^1)$; (2) $\eta_n^0$ and $\eta_n^1$ do not necessarily equal 1, and they do not necessarily equal each other. Before entering the oscillation, the vehicle is in the state with $\eta_n(t) = \eta_n^0$. After exiting the oscillation, the vehicle changes into another state with $\eta_n(t) = \eta_n^1$.

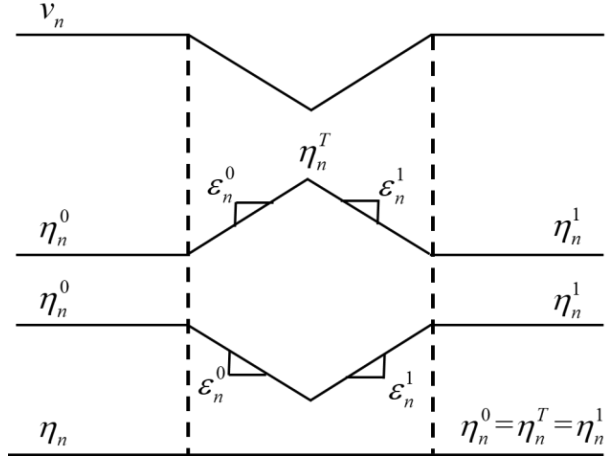

Figure 2. Variation patterns of $\eta_n(t)$. The dashed line separates the traffic oscillation from the equilibrium states.

However, the NGSIM US101 trajectory data is very short, about 40s for each trajectory, thus their information is pretty limited. In the following section, longer car following trajectories data will be investigated by analyzing the $\tilde{\tau}_n(t)$ time series. Since $\tilde{\tau}_n(t) = \eta_n(t)\tau$, the variation of $\tilde{\tau}_n(t)$ should also exhibit the same pattern as that of $\eta_n(t)$.

## 3 Experimental data analysis

*3.1. Data description*

In this section, the stationary states of the 25-car-platoon experiments (Jiang et al., 2014, 2015) will be extracted and analyzed. The experiments were carried out on a 3.2 km road in Hefei, China. In the experiment, the leading vehicle is required to drive at certain pre-determined constant speeds, see Figure 3 for some examples. Other drivers are required to drive as normally. Overtaking and lane-changing is prohibited during the experiment. The speed and location data are collected by the high-precision GPS devices every 0.1s.

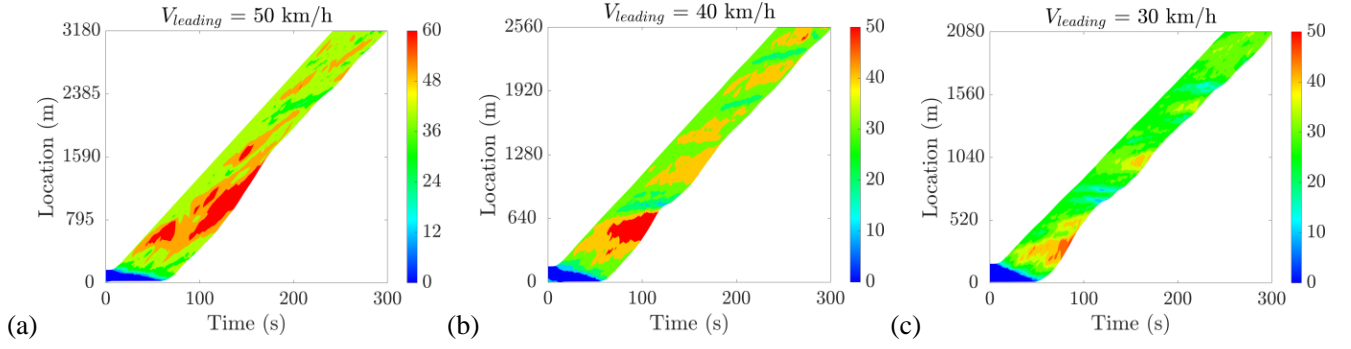

Figure 3. Some experimental spatiotemporal diagrams. From (a-c), the leading car moves with $v_{leading}$= 50, 40, 30 km/h, respectively.

*3.2 Some observed features of $\tilde{\tau}_n(t)$ time series*

To calculate the $\tilde{\tau}_n(t)$ time series, firstly $\tau$ and $w$ are calibrated. The Genetic algorithm is adopted and the parameters are optimized to minimize the Root Mean Square Percentage Error (*RMPSE*)

$$RMPSE = \frac{1}{M}\sqrt{\sum_{m=0}^{M}\left(\frac{d_n(mdt) - d_n^{Newell}(mdt)}{d_n(mdt)}\right)^2} \qquad (6)$$

where the length of time series is *Mdt*, with *dt* =0.1 s; *t = mdt, m* = 0, 1,…, *M*. $d_n(t) = x_{n-1}(t) - x_n(t)$ is the spacing from the follower's trajectory to the leader's trajectory. $d_n^{Newell}(t) = x_{n-1}(t) - x_n^{Newell}(t)$ is the spacing from the follower's Newell trajectory to the leader's trajectory.

Then the time series of $\tilde{\tau}_n(t)$ is extracted. We examine the time series of $\tilde{\tau}_n(t)$ for all trajectories and obtain the following remarks:

(i) No matter the speed of leading vehicle oscillates significantly (Figure 4) or slightly (Figure 5), $\tilde{\tau}_n(t)$ might change significantly.

(ii) A follower's $\tilde{\tau}_n(t)$ can vary from run to run even the leader travels at the same stable speed; see plot (a) vs. (b), (c) vs. (d) in Figure 5 for followers following the same leader.

(iii) Sometimes, even if the leader speed fluctuates significantly, the followers can keep a nearly constant value of $\tilde{\tau}_n(t)$ (see Figure 6).

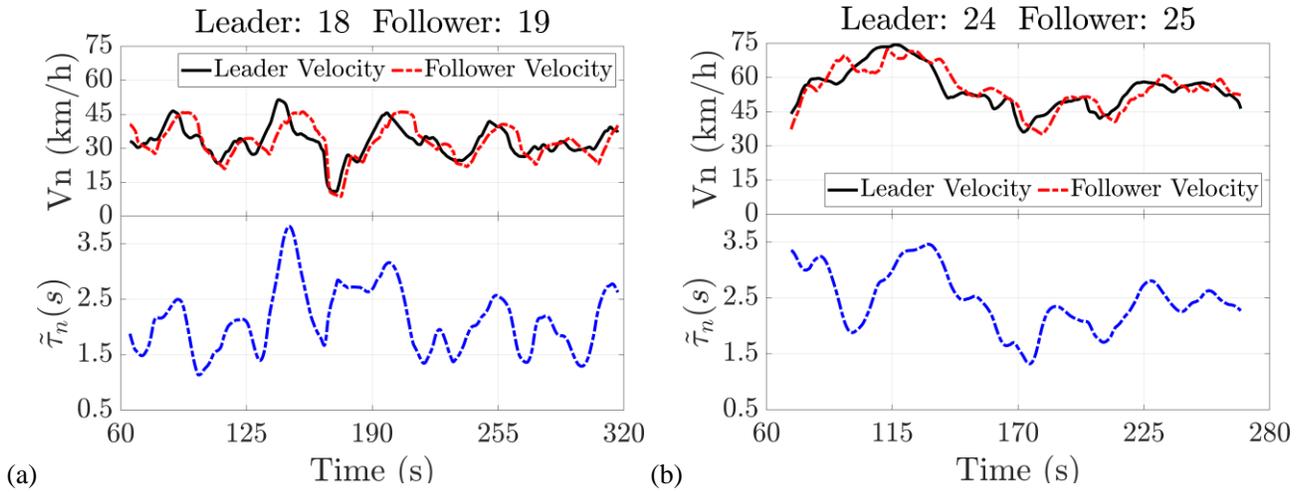

Figure 4. $\tilde{\tau}_n(t)$, $v_n(t)$ and $v_{n-1}(t)$ time series of the leader-follower pairs, in which speed of the leader remarkably fluctuates. The standard deviation of the leader's speed is (a) 1.96 m/s; (b) 2.49 m/s. The standard deviation of the follower's $\tilde{\tau}_n(t)$ is (a) 0.59 s; (b) 0.52 s.

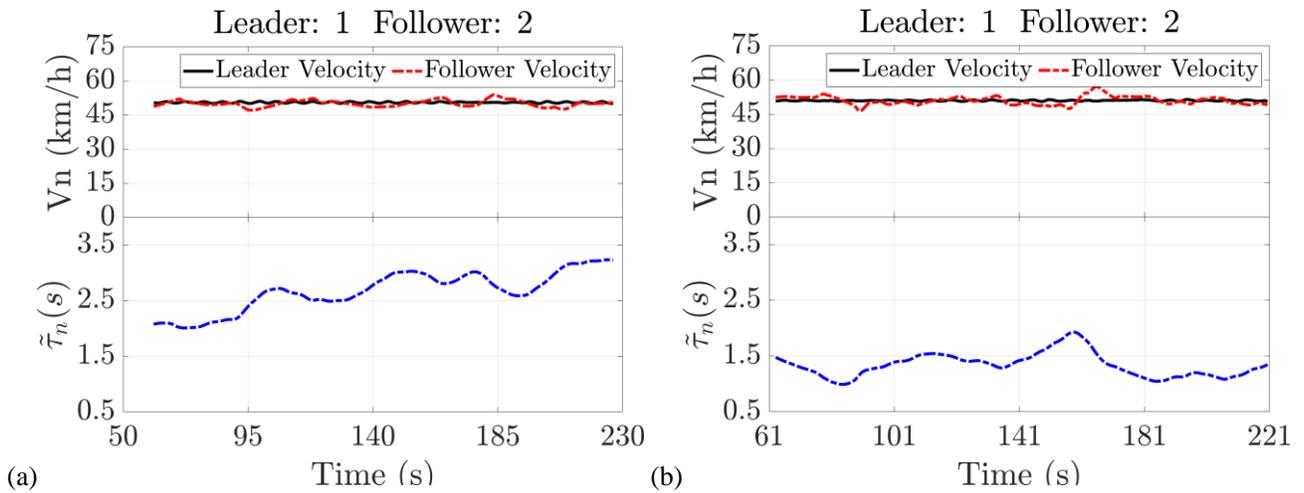

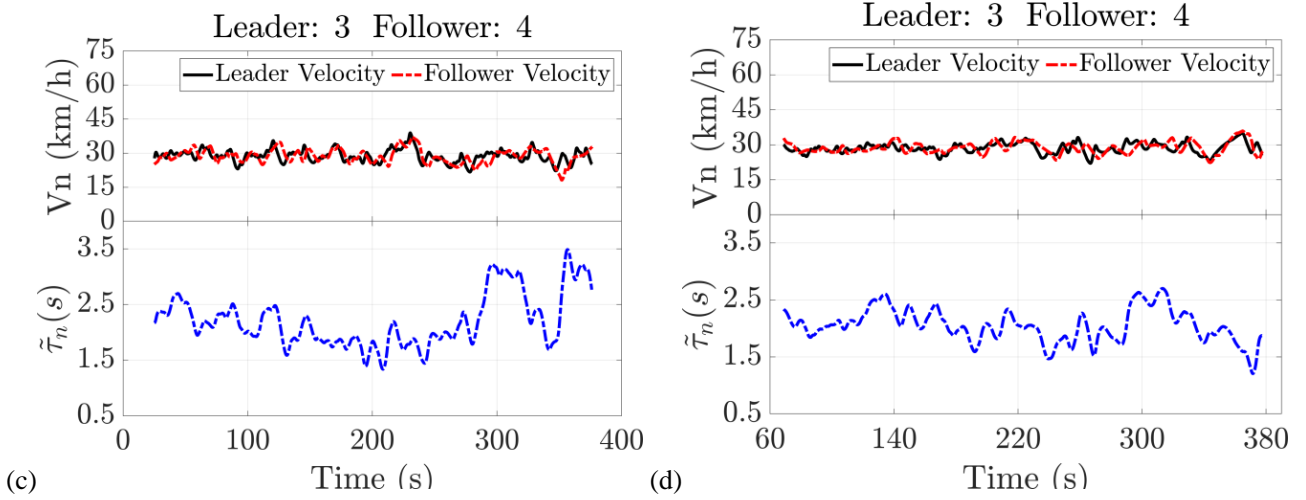

Figure 5. $\tilde{\tau}_n(t)$, $v_n(t)$ and $v_{n-1}(t)$ time series of the leader-follower pairs, in which speed of the leader only slightly fluctuates. The standard deviation of the leader's speed is 0.07, 0.06, 0.76, 0.62 m/s in (a)-(d); The standard deviation of the follower's $\tilde{\tau}_n(t)$ is 0.35, 0.22, 0.48, 0.29 s in (a)-(d).

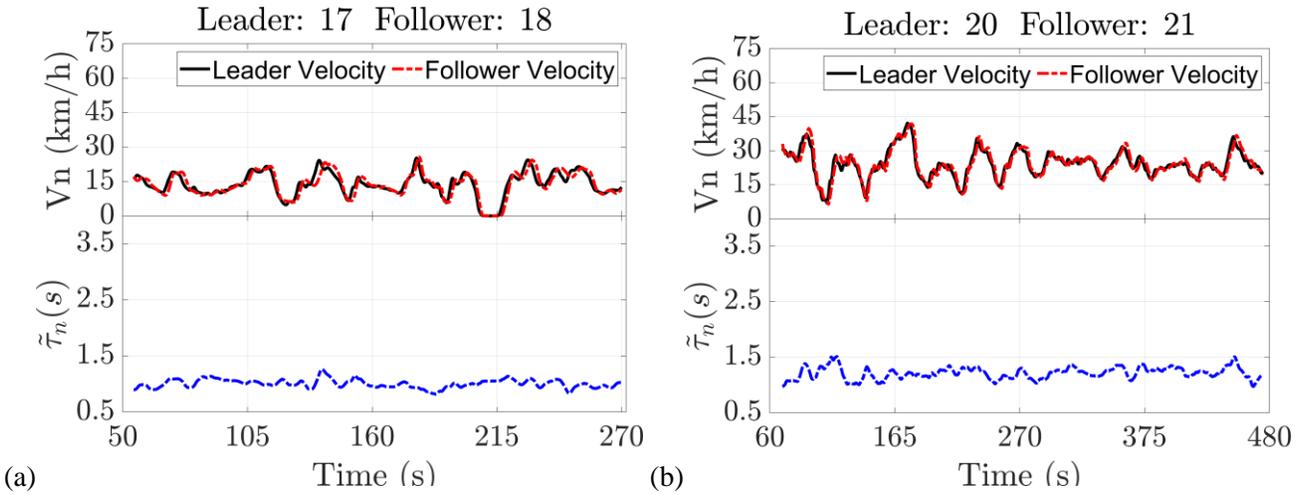

Figure 6. $\tilde{\tau}_n(t)$, $v_n(t)$ and $v_{n-1}(t)$ time series of the leader-follower pairs. The standard deviation of the leaders' speed is (a) 1.43 m/s; (b) 1.68 m/s. The standard deviation of the follower's $\tilde{\tau}_n(t)$ is (a) 0.07 s; (b) 0.11 s.

Now we study the time series of the wave travel time changing rate $\xi_n(t)=d\tilde{\tau}_n(t)/dt$. Since GPS devices collected the speed and location data every 0.1s, we calculate the changing rate via $\xi_n(t)=(\tilde{\tau}_n(t)-\tilde{\tau}_n(t-dt))/dt$, with $dt= 0.1$s. Figure 7 shows three examples of time series of $\xi_n(t)$, which oscillates around zero.

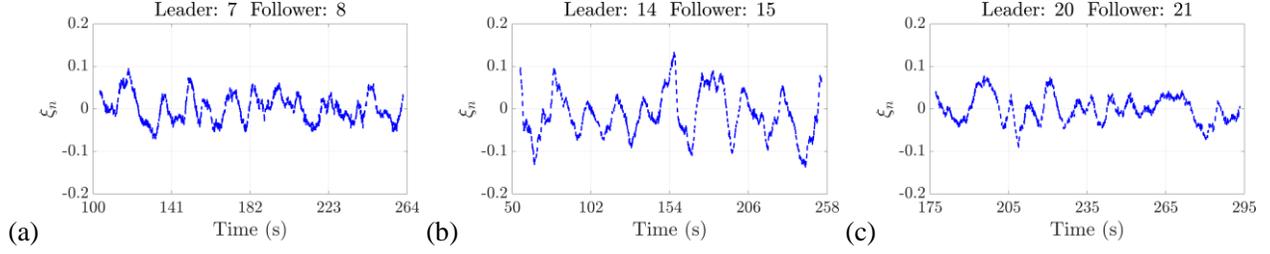

Figure 7. The time series of $\xi_n(t)$.

To study feature of the time series of $\xi_n(t)$, we have conducted the Augmented Dickey-Fuller (ADF) test. The results show that 99.04% (618 trajectories of total 624 trajectories) of $\xi_n(t)$ time series follows the mean reversion process (Chaudhuri and Wu, 2003).

Therefore, we can use the mean reversion model to capture $\xi_n(t)$. To this end, the Vasicek model was applied, which was originally used to describe the evolution of the instantaneous interest rate and widely applied to simulate the mean reversion process (Vasicek, 1977).

$$d\xi_n = \alpha(\mu-\xi_n)dt + \sigma dW_t \tag{7}$$

where $W_t$ is a standard Wiener process; the evolution of $\xi_n(t)$ will evolve around the mean level $\mu$ in the long run; $\alpha$ is the mean reversion speed, which characterizes the rate of evolution in recovering $\mu$; $\sigma$ is the volatility rate that measures the amplitude of randomness in the system. Higher $\sigma$ implies larger randomness.

To calibrate Equation (7), The maximum likelihood estimation (Tang and Chen, 2009) is applied. To this end, the solution of $\xi_n(t)$ is given as follows (Karatzas and Shreve, 1991),

$$\xi_n(t) = \xi_n(0)e^{-\alpha t} + \mu(1-e^{-\alpha t}) + \sigma \int_0^t e^{-\alpha s} dW(s) \tag{8}$$

which follows the normal distribution with the mean and variance:

$$\mu_n^\xi \triangleq E(\xi_n(t)) = \xi_n(0)e^{-\alpha t} + \mu(1-e^{-\alpha t}) \tag{9}$$

$$\left(\sigma_n^\xi\right)^2 \triangleq \operatorname{Var}\left(\xi_n(t)\right) = \frac{\sigma^2}{2\alpha}\left(1 - e^{-2\alpha t}\right) \tag{10}$$

Then the conditional mean and variance of $\xi_n(t)$ given $\xi_n(t - dt)$ are

$$\operatorname{E}\left(\xi_n(t) \mid \xi_n(t - dt)\right) = \xi_n(t - dt) e^{-\alpha dt} + \mu\left(1 - e^{-\alpha dt}\right) \tag{11}$$

$$\operatorname{Var}\left(\xi_n(t) \mid \xi_n(t - dt)\right) = \frac{\sigma^2}{2\alpha}\left(1 - e^{-2\alpha dt}\right) \tag{12}$$

Denote the length of time series of $\xi_n$ by $Mdt$ and $t = mdt$, $m = 0, 1, \ldots, M$, the time series of $\xi_n(t)$ is rewritten as $\xi_n(mdt)$ and further simplified as $\xi_n(m)$. The likelihood function is

$$L = \phi\left(\sigma^{-1}\sqrt{2\alpha}\left(\xi_n(0) - \mu\right)\right) \prod_{m=1}^{M} \phi\left(\frac{\xi_n(m) - \left(\xi_n(m-1) e^{-\alpha dt} + \mu\left(1 - e^{-\alpha dt}\right)\right)}{\sqrt{\frac{\sigma^2}{2\alpha}\left(1 - e^{-2\alpha dt}\right)}}\right) \tag{13}$$

where $\phi$ is the density function of the standard normal distribution. Therefore the maximum likelihood estimators is obtained by

$$\hat{\alpha} = -\frac{1}{dt}\log(\eta_1),\ \hat{\mu} = \frac{M^{-1}\sum_{m=1}^{M}\left(\xi_n(m) - \eta_1 \xi_n(m-1)\right)}{1 - \eta_1},\ \hat{\sigma} = \sqrt{\frac{2\hat{\alpha}\eta_2}{1 - \eta_1^2}} \tag{14}$$

where

$$\eta_1 = \frac{\sum_{m=1}^{M}\xi_n(m-1)\xi_n(m) - M^{-1}\sum_{m=1}^{M}\xi_n(m)\sum_{m=1}^{M}\xi_n(m-1)}{\sum_{m=1}^{M}\xi_n^2(m-1) - M^{-1}\left(\sum_{m=1}^{M}\xi_n(m-1)\right)^2} \tag{15}$$

$$\eta_2 = M^{-1}\sum_{m=1}^{M}\left(\xi_n(m) - \eta_1 \xi_n(m-1) - \hat{\mu}(1 - \eta_1)\right)^2 \tag{16}$$

The statistical results of calibrated parameters are shown in Table 1, which shows that the estimation results have high statistical significance. It indicates that the Vasicek model delivers satisfactory performance in simulating the $\xi_n(t)$ time series.

Table 1. Statistics of $\mu$, $\alpha$ and $\sigma$.

|   | Simple size | Median | Mean | Std | Mean (absolute t-statistics) |
|---|---|---|---|---|---|
| $\alpha$ | 624 | 0.265 | 0.432 | 0.511 | 73.6 |
| $\mu$ | 624 | 0.001 | 0.001 | 0.008 | 5.1 |
| $\sigma$ | 624 | 0.035 | 0.042 | 0.029 | 859.3 |

## 4 New Experimental data analysis

In the 25 car-platoon, the oscillation has not yet fully developed[1] due to the limit platoon length. Therefore, we perform a new set of experiments to mimic the fully developed oscillations and study evolution of $\tilde{\tau}_n(t)$ in the fully developed oscillations. In the experiments, we use a 5-car-platoon. The experiments were carried out on a remote suburb road with length about 4 km. There is no traffic light on the road and no other vehicle intervening the experiment. The experiment instructions are as follows. Initially the platoon stops at one end of the road. For the leading car, the driver is asked to

(i) accelerate to 40 km/h, then maintain the 40 km/h speed for 30 s to 1 min.
(ii) randomly choose to decelerate to 10 km/h or complete stop with equal probability.
(iii) maintain the 10 km/h speed or stop for 5-10 s, then accelerate back to 40 km/h.
(iv) maintain the 40 km/h speed for 2-3 min.
(v) repeat the deceleration and acceleration process.
(vi) maintain the 40 km/h speed until arriving at the other end of the road and decelerate to stop.
(vii) Make a U-turn and prepare for the next round of experiment.

---

[1] Here fully developed oscillations are defined as: oscillations are well separated from each other. Between two consecutive oscillations, traffic flow is nearly stationary.

For following cars, the drivers are asked to drive as normally. Overtaking is not allowed. High precise GPS data with interval of 0.1s were collected. Figure 8 shows typical trajectories and speed time series of the five cars in one round of experiment.

To study evolution of $\tilde{\tau}_n(t)$ related to each oscillation, we classify each round into two intervals. Each interval includes one oscillation and the following stationary traffic flow. We have extracted 40 intervals. In 19 intervals, the leading car has stopped. In 21 intervals, the leading car decelerated to 10 km/h. Figure 8 shows one example of $\tilde{\tau}_n(t)$ time series for these two types of oscillations.

We calculate the $\xi_n(t)$ time series from the $\tilde{\tau}_n(t)$ time series. We have four drivers, and 40 intervals. Therefore, we have totally 160 $\xi_n(t)$ time series. We perform the ADF test, which indicates that all the 160 $\xi_n(t)$ time series also follow the mean reversion process. Therefore, the Vasicek model is calibrated. The results are given in Table 2, which also shows that the estimation results have high statistical significance.

Table 2. Statistics of $\mu$, $\alpha$ and $\sigma$.

|   | Simple size | Median | Mean | Std | *Mean*(*absolute t-statistics*) |
|---|---|---|---|---|---|
| $\alpha$ | 160 | 0.265 | 0.337 | 0.223 | 49.2 |
| $\mu$ | 160 | 0.001 | 0.002 | 0.006 | 4.3 |
| $\sigma$ | 160 | 0.028 | 0.029 | 0.012 | 632.1 |

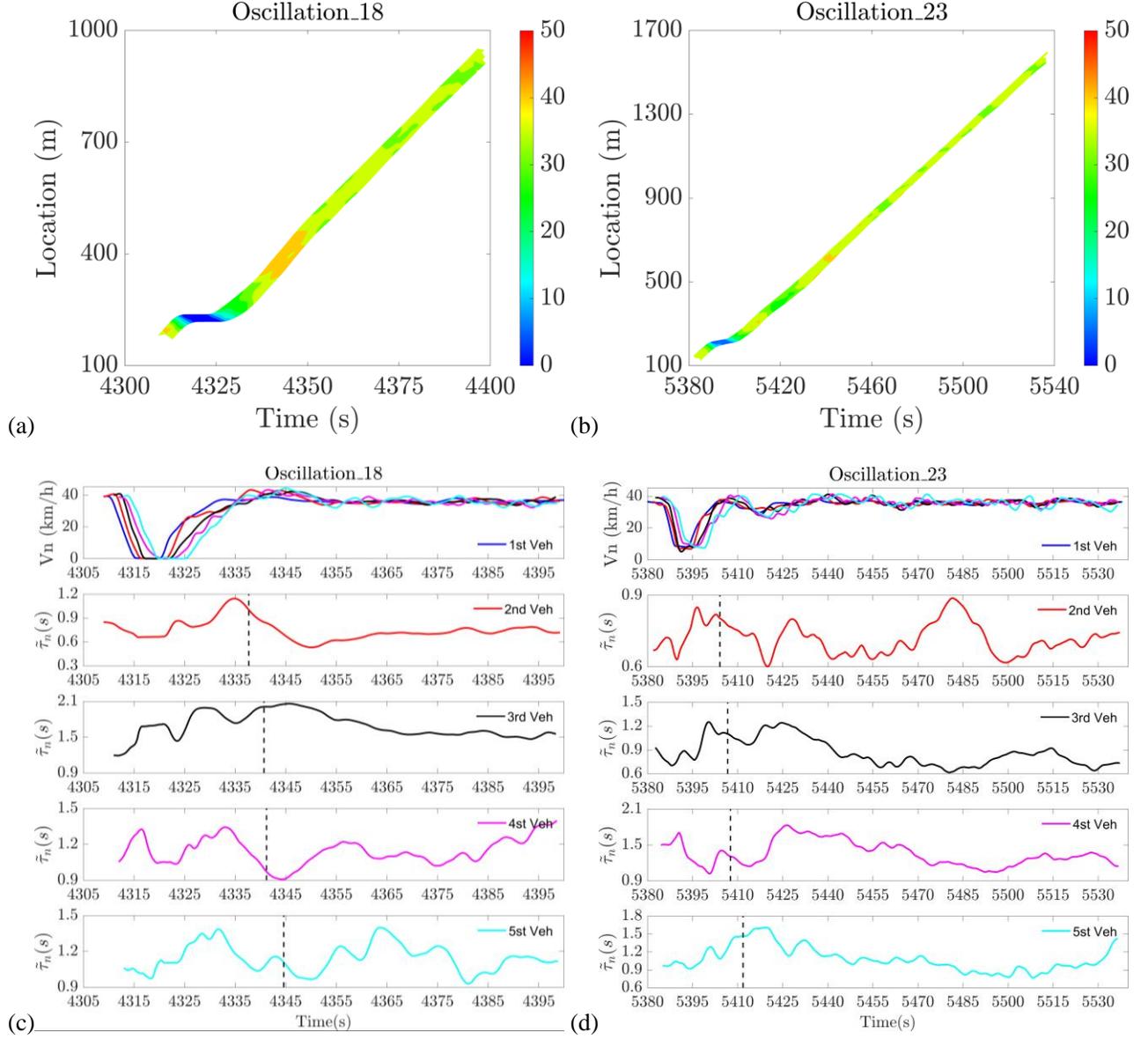

Figure 8. (a, b) Samples of the vehicle trajectories with relocated position that starts from zero and (c-d) the speeds and $\tilde{\tau}_n(t)$ corresponding to (a, b), respectively. The dashed line roughly separates the oscillation from the stationary traffic flow.

## 5 The new car following model

Based on above analysis, a simple car following model is proposed.

$$v_n^{\text{free}}(t+\tau) = \min\left(v_{\max}, v_n(t) + a_n(t)\tau\right) \tag{17}$$

$$x_n(t+\tau) = \min\left(x_n(t) + v_n^{\text{free}}(t+\tau)\tau, x_{n-1}(t) - w\tilde{\tau}_n(t)\right) \tag{18}$$

The wave travel time $\tilde{\tau}_n(t)$ is updated by the following equations (see the derivations in the Appendix).

$$\tilde{\tau}_n(t+\tau) = \left[\tilde{\tau}_n(t) + \varepsilon_n(t)\right]_{\tilde{\tau}_{min}}^{\tilde{\tau}_{max}} \tag{19}$$

where $\varepsilon_n(t)$ follows the normal distribution with mean $\mu_n^\varepsilon = 0$ and variance $\sigma_n^\varepsilon = \tau\tilde{\sigma}$, where $\tilde{\sigma} = \dfrac{\sigma}{\sqrt{2\alpha}}$. It should be noted that $\tilde{\tau}_n(t) \geq \tilde{\tau}_{min} = L_{veh}/w$ is imposed to avoid accident and $\tilde{\tau}_n(t)$ is bounded up by $\tilde{\tau}_{max}$ to restrict the influence of the leading car when the gap is large enough. The acceleration $a_n(t)$ of vehicle $n$ is defined as follows.

$$a_n(t) = a\left(1 - \dfrac{v_n(t)}{v_{max}}\right) \tag{20}$$

**Theorem:** Assuming there is a car following platoon with $N+1$ vehicles (0 to $N$) in a lane, the leading vehicle (with number 0) moves with a constant speed $\bar{v}$, then speed standard deviations of vehicle $n$ at time $t$ can be approximated by $\sqrt{n}\sigma_n^\varepsilon$.

Proof: According to the new model, the movement of vehicles can be approximated by

$$x_n(t+\tau) = x_{n-1}(t) - w\tilde{\tau}_n(t) \tag{21}$$

$$x_n(t) = x_{n-1}(t-\tau) - w\tilde{\tau}_n(t-\tau) \tag{22}$$

thus

$$v_n(t) = v_{n-1}(t-\tau) - \dfrac{w}{\tau}\left(\tilde{\tau}_n(t) - \tilde{\tau}_n(t-\tau)\right) \approx v_{n-1}(t-\tau) - \dfrac{w}{\tau}\varepsilon_n(t-\tau), \tag{23}$$

which yields

$$v_n(t) \approx \bar{v} - \sum_{i=1}^{n} \dfrac{w}{\tau} \varepsilon_n\left(t - (n-i+1)\tau\right) \tag{24}$$

Since $\varepsilon_n$ is independent among vehicles,

$$\mathrm{Var}(v_n(t)) \approx \dfrac{w^2}{\tau^2} \sum_{i=1}^{n} \mathrm{Var}\left(\varepsilon_n\left(t-(n-i+1)\tau\right)\right) = n\left(\sigma_n^\varepsilon\right)^2 \tag{25}$$

Thus,

$$\text{STD}(v_n(t)) \approx \sqrt{n}\sigma_n^{\varepsilon} \qquad (26)$$

Figure 9 compares the model simulation results with the analytical results calculated by Equation (26). It can be seen that Equation (26) can predict the speed standard deviations of the vehicles along the platoon quite well.

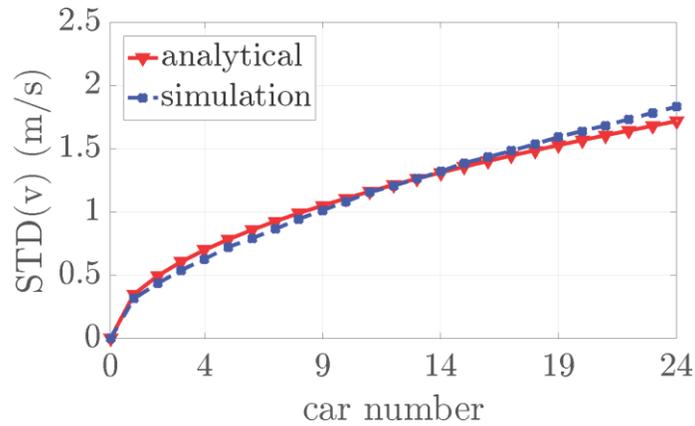

Figure 9. The standard deviation of speed. The car number 0 is the leading car and the leading vehicle runs at a constant speed of 40 km/h. "analytical" is the results calculated by Equation (26). "simulation" is the model simulation results. The parameters are taken from Table 3.

## 6 The Simulations

In this section, several different simulations are conducted to investigate performance of the new model. Firstly, the platoon oscillation evolutions are simulated to examine whether macroscopic characteristics of car-following platoon, including the concave growth patterns of speed standard deviations and the platoon length, can be quantitatively reproduced or not. Secondly, the bottleneck road system is simulated to examine whether the empirical observed congested pattern of traffic flow can be simulated or not. Finally, vehicle trajectories are simulated to investigate the new model's ability to simulate the microscopic aspects of car following behaviors.

*6.1 Platoon oscillation simulations*

Here we show that the new model can quantitatively reproduce the concave growth of platoon oscillations. To this end, the model parameter values are calibrated by the Genetic Algorithm to minimize the average *RMPSE* of the simulated and real speed standard deviations ($\sigma_{v_n}$) of vehicles and platoon length ($l_p$), i.e.

$$RMPSE = \frac{1}{2(N-1)}\sqrt{\sum_{n=2}^{N}\left(\frac{\sigma_{v_n}^{\text{sim}} - \sigma_{v_n}^{\text{emp}}}{\sigma_{v_n}^{\text{emp}}}\right)^2} + \frac{1}{2}\left|\frac{l_p^{\text{sim}} - l_p^{\text{emp}}}{l_p^{\text{emp}}}\right| \qquad (27)$$

where *N*-1=24 is the number of following vehicles. The results are given in Table 3. In the simulation, the vehicle length is set as $L_{\text{veh}} = 5$ m.

Table 3. Parameter values of the new model.

| Parameter | $v_{\max}$ | $a$ | $\tau$ | $\tilde{\sigma}$ | $s_0$ | $\tilde{\tau}_{\max}$ |
|---|---|---|---|---|---|---|
| Value | 80 | 0.5 | 1.1 | 0.055 | 2.0 | 2.5 |
| Unit | km/h | m/s$^{-2}$ | s | s | m | s |

The simulation results and the corresponding experimental ones are shown in Figure 10. It can be seen that: (1) the simulated speed standard deviation increases in a concave way along the platoon and agrees with the experimental results quite well; (2) the simulated spatiotemporal patterns reproduce the stripe structures as appearing in the experimental ones (Figure 3); (3) the platoon length also agrees with the experimental results very well.

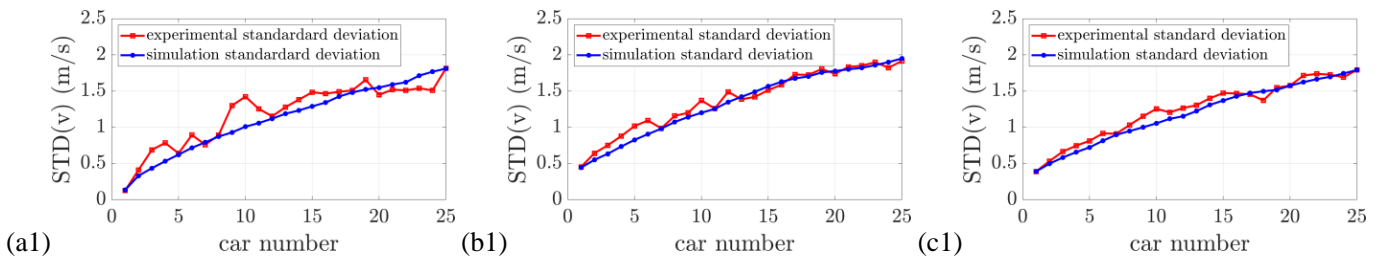

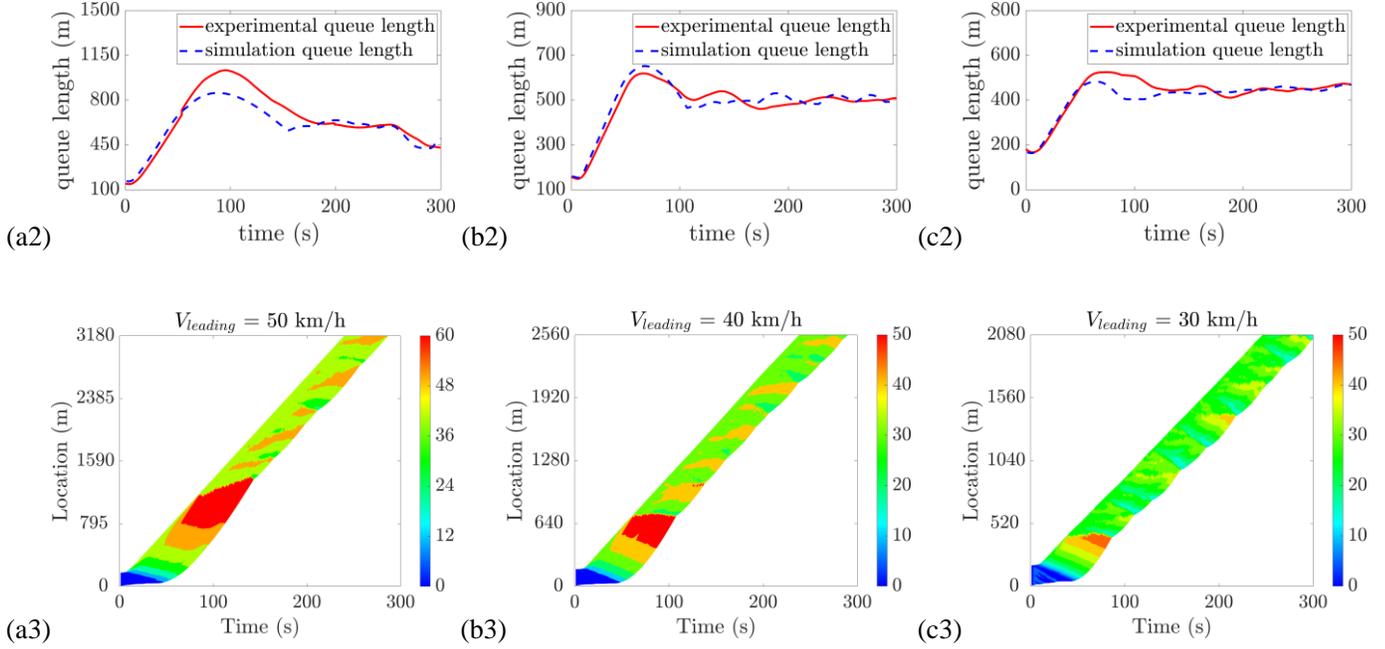

Figure 10. The standard deviation of speed (a1) - (c1), the platoon length (a2) - (c2) and the simulated spatiotemporal diagrams (a3) - (c3). The car number 1 is the leading car. From (a1) - (c1), (a2) -(c2), (a3) - (c3) the leading car moves with $v_{\text{leading}}$=50, 40 and 30 km/h, respectively.

## 6.2. Empirical congested pattern simulation

There are two different congested patterns in the empirical spatiotemporal traffic flow patterns in NGSIM data, which were collected on a 640 m segment on southbound US 101 in Los Angeles, CA, on June 15th, 2005 between 7:50 a.m. and 8:35 a.m., see Figure 11 (a) and (b). Figure 11 (a) shows that the jams emerge from the congested traffic flow in the bottleneck region and then propagate upstream along the road, thus the congested traffic flow is unstable. In Figure 11 (b), no jams emerge from the congested traffic flow, i.e. the traffic flow is stable. Chen et al. (2012a) argued that the rubbernecking caused by the clean-up work on Lane 1 between 7:50 a.m. and 8:05 a.m. is the most likely cause for these congested traffic flow in this segment. The rubbernecking zone is located at [320, 420] m, where the road length $L_{\text{road}} = 640$ m.

To simulate the rubbernecking bottleneck, the following rule is adopted. When vehicles enter the rubbernecking zone, the drivers have a probability $p_{rub}$ to rubberneck which will lead them to decelerate with deceleration $d_{rub}$ for $h_{rub}$ second. Rubbernecking occurs at most once for the drivers in the bottleneck region. During the simulation, the bottleneck is set to be located at [$0.8L_{road}$, $0.8L_{road}+L_{bottleneck}$] with the length $L_{bottleneck}$=100 m. The road is initially assumed to be filled with cars that are uniformly distributed with density $k$ and speed $v_{max}$. For the leading car going beyond $L_{road}$, it will be removed from the simulation. Figure 11 (c) and (d) show the simulation results of the new model. The stable and unstable congested patterns can be successfully simulated by the new model.

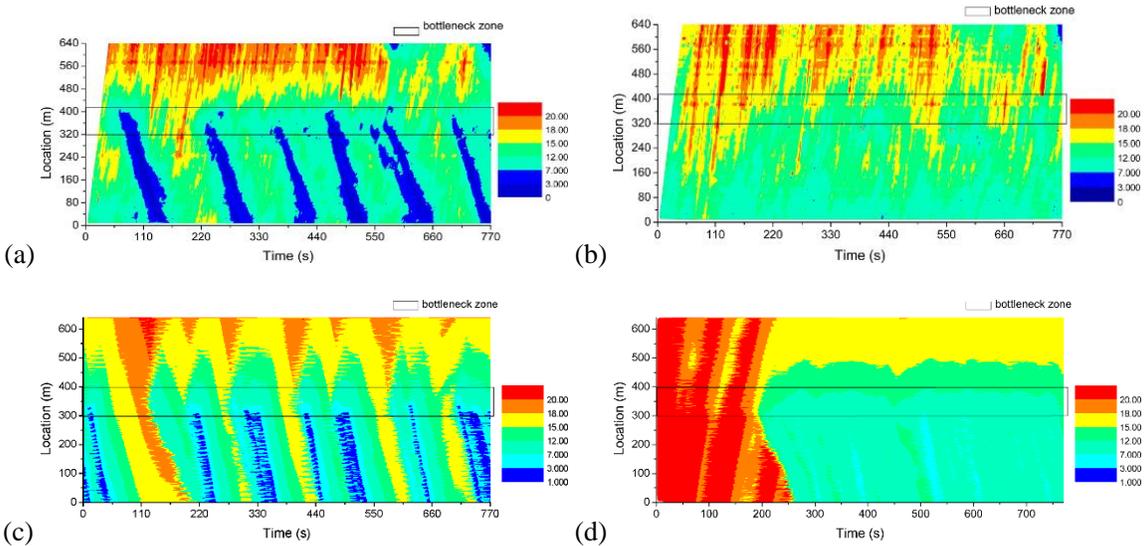

Figure 11. Empirical spatiotemporal patterns of the velocity on the Lane 1 (a) and Lane 5 (b) from the NGSIM trajectory dataset collected on a 640 m segment on southbound US 101. The simulation of congested traffic flow on an open road with a rubberneck bottleneck with (c) $p_{rub}$ = 0.05, $d_{rub}$ = 2.25 m/s$^2$, $h_{rub}$ =6.0 s and (d) $p_{rub}$ = 0.5, $d_{rub}$ = 0.6 m/s$^2$, $h_{rub}$ = 2.0 s. The color bar indicates speed (unit: m/s). The parameter values are: $v_{max}$=80 km/h, $a$=1.0 m/s$^2$, $\tau$=1.1 s, $\tilde{\sigma}$ =0.055 s, $s_0$=2.0 m and $\tilde{\tau}_{max}$ =2.5 s.

## 6.3. Vehicle trajectory simulations

In this section, the new model is used to calibrate the trajectory of the 5-car-platoon experiments reported in section 4. In the calibration process, we use GA algorithm and the trajectory of each paired leading-following vehicle is used.

In each run, RMPSE between real spacing and simulated spacing is calculated.

$$RMPSE_r = \frac{1}{M}\sqrt{\sum_{m=0}^{M}\left(\frac{d_n(t)-d_n^{sim}(t)}{d_n(t)}\right)^2} \qquad (28)$$

where $d_n^{sim}(t) = x_{n-1}(t) - x_n^{sim}(t)$ is the spacing from the follower's simulated trajectory to the leader's trajectory.

We perform R = 200 runs, minimizing the Average RMPSE (ARMPSE)

$$ARMPSE = \frac{1}{R}\sum_{r=1}^{R}RMPSE_r \qquad (29)$$

Figure 12 shows the distribution of ARMPSE and the statistic of ARMPSE over 160 trajectories. One can see that the errors are rather small. Figure 13 shows two examples of trajectory simulation. The real data and the mean value of simulated spacing and speed are plotted as blue line and red line, respectively. The corresponding 5% - 95% probability band is shown as a gray area. The difference between simulation results and real data is small, which demonstrates that this model can reproduce the microscopic characteristics in car following process well.

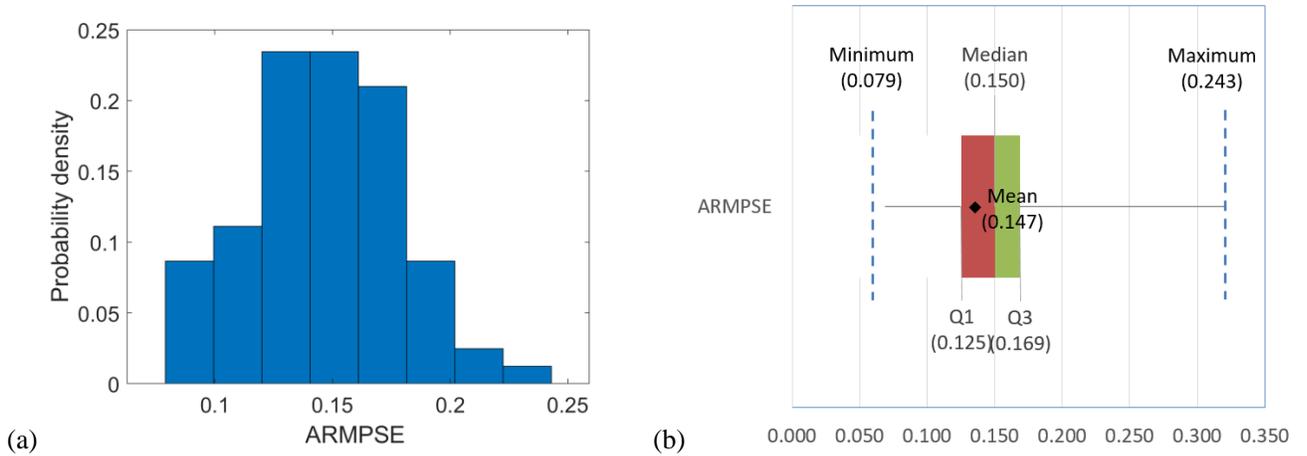

Figure 12. The distribution of the ARMPSE (a) and statistics of ARMPSE (b).

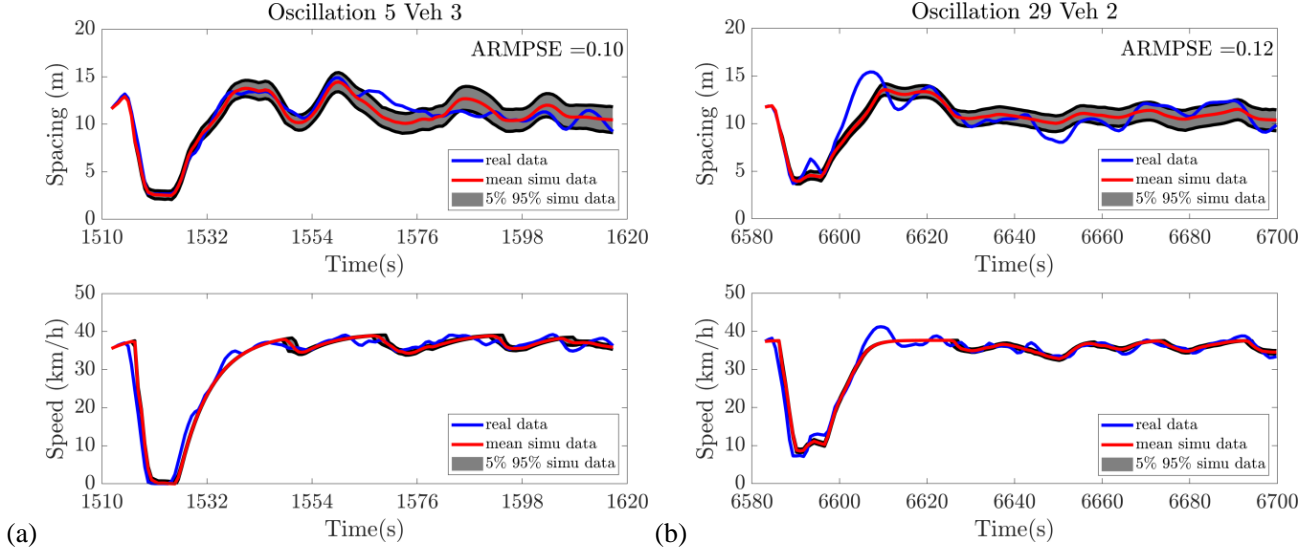

Figure 13. The spacings and speeds of the simulated vehicle. In (a), the parameters are: $v_{max}$=40 km/h, $a$=1.51 m/s$^2$, $\tau$=0.7 s, $\tilde{\sigma}$=0.01 s, $s_0$=2.51 m and $\tilde{\tau}_{max}$=1.8 s. In (b), the parameters are: $v_{max}$=38 km/h, $a$=3.73 m/s$^2$, $\tau$=0.8 s, $\tilde{\sigma}$=0.01 s, $s_0$=1.82 m and $\tilde{\tau}_{max}$=1.62 s.

## 7. Discussion

In a recent study (Tian et al., 2019), it has been found that the time series of acceleration $a_n(t)$ and the speed difference $\Delta v_n(t-\tau)$ between a car $n$ and its front car $n$-1 exhibit striking similarities. Here $\tau$ is a time delay. After calculating $\lambda$ through linear regression of $a_n(t) = \lambda \Delta v_n(t-\tau)$, it is shown that the residual time series $\zeta(t) = a_n(t) - \lambda \Delta v_n(t-\tau)$ also follows the mean reversion process. A Stochastic Speed Adaptation (SSA) car-following model has been proposed based on that finding, which well reproduces the 25-vehicle-experimental results.

Next, we study the relationship between the two findings. To this end, we extract $\tilde{\tau}_n(t)$ of follower from the simulated trajectories of two successive vehicles in the SSA model, using the same parameters as in Tian et al. (2019). Then the time series of $\xi_n(t)$ is studied. Figure 14 shows the simulation results. The ADF test indicates that the time series of $\xi_n(t)$ generated by the SSA model also follows the mean reversion process and can be simulated by Equation (7). This indicates that the stochastic factors might exhibit common feature (i.e., mean reversion process) no matter from perspective of wave travel time or from perspective of speed adaptation.

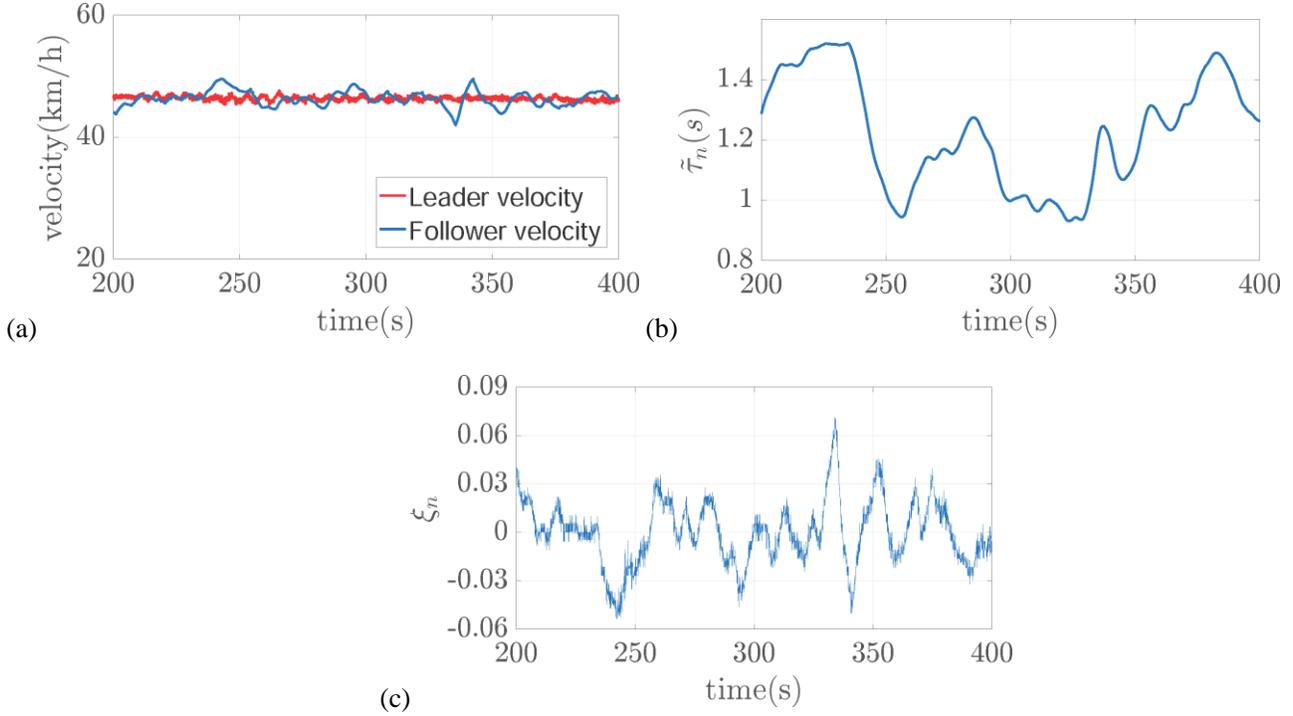

Figure 14. The simulation results of the SSA model. (a) is the velocity of the leader ($v_{n-1}$) and the follower ($v_n$) and the spacing of the follower. (b) the $\tilde{\tau}_n(t)$ time series of the follower. (c) the $\xi_n(t)$ time series of the follower, in which $\alpha = 0.71$ and $\sigma = 0.022$s.

## 8 Conclusions

This paper conducts a detailed analysis on the car-following stochasticity following the framework of Laval and Leclercq (2010) by measuring the wave travel time series $\tilde{\tau}_n(t)$ from leader to the follower. To this end, two sets of field experimental trajectory data are analyzed. In the first set of field experiment, the oscillations form and grow spontaneously along the vehicles. In the second set of field experiment, the leader vehicle moves mimicking the fully developed oscillations. The most significant findings are: (i) the wave travel time does not follow a consistent pattern no matter the speed of the leading vehicle significantly fluctuates or is nearly constant; (ii) the time series $\xi_n(t)=d\tilde{\tau}_n(t)/dt$ follows a mean reversion process, no matter the oscillations fully developed or not. Based on these findings, a new stochastic Newell type model is proposed. Simulations show that the new model is able to reproduce the concave growth pattern of traffic oscillations and the stable and unstable congested traffic flow patterns observed in the NGSIM US101 dataset. Moreover, the microscopic characteristic of car following can be well simulated. Finally, a comparison with previous study demonstrates that stochastic factors might exhibit common feature (i.e., mean reversion process) from different perspectives, either from perspective of wave travel time or from perspective of speed adaptation.

We would like to mention that in both sets of experiment, the platoon length is limited. In particular, in the second set of experiment, the platoon has only 5 cars and the cruising speed is fixed as 40 km/h. In the future work, more experimental data are needed for further investigations.

## Appendix:

Since $\frac{d\tilde{\tau}_n}{dt} = \xi_n$, the distribution of $\varepsilon_n(t) = \tilde{\tau}_n(t+\tau) - \tilde{\tau}_n(t) = \int_t^{t+\tau} \xi_n(s)ds$ follows the normal distribution with the mean and variance:

$$\mu_n^\varepsilon \triangleq \mathrm{E}\left( \int_t^{t+\tau} \left( \xi_n(0)e^{-\alpha s} + \mu(1-e^{-\alpha s}) + \sigma \int_0^s e^{-\alpha u} dW_n(u) \right) ds \right)$$
$$= \mu\tau - \left(e^{-\alpha t} - e^{-\alpha(t+\tau)}\right)\left(\mu - \xi_n(0)\right)/\alpha \tag{A1}$$

$$\left(\sigma_n^\varepsilon\right)^2 \triangleq \sigma^2 E\left( \int_t^{t+\tau} \int_0^s e^{-\alpha u} dW_n(u) ds \right)^2 \tag{A2}$$

Denote $Y(t) \triangleq \int_0^t e^{-\alpha u} dW_n(u)$, Ito's rule (Karatzas and Shreve, 1991,) implies that

$$\int_t^{t+\tau} Y(s)ds = (t+\tau)Y(t+\tau) - tY(t) - \int_t^{t+\tau} se^{-\alpha s} dW_n(s) \tag{A3}$$

Then we have,

$$\left(\sigma_n^\varepsilon\right)^2 = \sigma^2 E\left[(t+\tau)Y(t+\tau) - tY(t) - \int_t^{t+\tau} se^{-\alpha s} dW_n(s)\right]^2$$

$$= \sigma^2 \begin{pmatrix} (t+\tau)^2 E(Y(t+\tau))^2 + t^2 E\left(\int_0^t e^{-\alpha s} dW_n(s)\right)^2 \\ + E\left(\int_t^{t+\tau} se^{-\alpha s} dW_n(s)\right)^2 - 2(t+\tau)E\left(\int_0^{t+\tau} e^{-\alpha s} dW_n(s) \int_t^{t+\tau} se^{-\alpha s} dW_n(s)\right) \\ -2t(t+\tau)E\left(\int_0^{t+\tau} e^{-\alpha s} dW_n(s) \int_0^t e^{-\alpha s} dW_n(s)\right) + 2tE\left(\int_0^t e^{-\alpha s} dW_n(s) \int_t^{t+\tau} se^{-\alpha s} dW_n(s)\right) \end{pmatrix} \quad (A4)$$

Since the quadratic variation of Brownian motion ($W(t)$), denoted by $<W(t), W(t)>$, is time $t$, and $W^2(t) - t$ is a martingale, we have,

$$E\left(\int_t^{t+\tau} e^{-\alpha s} dW_n(s)\right)^2 = \int_t^{t+\tau} e^{-2\alpha s} ds \quad (A5)$$

$$E\left(\int_t^{t+\tau} se^{-\alpha s} dW_n(s)\right)^2 = \int_t^{t+\tau} s^2 e^{-2\alpha s} ds \quad (A6)$$

$$E\left(\int_t^{t+\tau} e^{-\alpha s} dW_n(s) \int_t^{t+\tau} se^{-\alpha s} dW_n(s)\right) = \int_t^{t+\tau} se^{-2\alpha s} ds \quad (A7)$$

$$E\left(\int_0^t e^{-\alpha s} dW_n(s) \int_t^{t+\tau} se^{-\alpha s} dW_n(s)\right) = 0 \quad (A8)$$

Hence,

$$\left(\sigma_n^\varepsilon\right)^2 = \sigma^2 \left(\frac{\tau^2}{2\alpha} - \frac{\tau e^{-2\alpha t}}{2\alpha^2} + \frac{e^{-2\alpha t} - e^{-2\alpha(t+\tau)}}{4\alpha^3}\right) \quad (A9)$$

Since $t \gg 1$ in the simulation and data analysis shows that $\mu$ approximates to zero, we can let

$$\mu_n^\varepsilon \approx 0 \quad (A10)$$

$$\sigma_n^\varepsilon \approx \frac{\tau\sigma}{\sqrt{2\alpha}} \quad (A11)$$

to simplify the model. Furthermore, since $\tilde{\tau}_n(t)$ is bounded by $\tilde{\tau}_{\min}$ and $\tilde{\tau}_{\max}$, therefore, $\tilde{\tau}_n(t)$ is updated by $\tilde{\tau}_n(t+\tau) = \left[\tilde{\tau}_n(t) + \varepsilon_n(t)\right]_{\tilde{\tau}_{\min}}^{\tilde{\tau}_{\max}}$ and $\varepsilon_n(t) \sim N\left(0, \frac{\tau\sigma}{\sqrt{2\alpha}}\right)$.

## Acknowledgements:

JFT was supported by the National Natural Science Foundation of China (Grant No. 71771168, 71671123). RJ was supported by the Natural Science Foundation of China (Grant Nos. 71931002, 71621001, 71631002).